# The role of the Allee effect in common-pool resource and its sustainability

**Authors:** Chengyi Tu [a, 1, *], Fabio Menegazzo [b, 1], Paolo D'Odorico [c], Samir Suweis [b, *]

[a] School of Economics and Management, Zhejiang Sci-Tech University; Hangzhou, 310018, China.

[b] Department of Physics and Astronomy "G. Galilei", University of Padua; Padova, 35131, Italy.

[c] Department of Environmental Science, Policy, and Management, University of California, Berkeley; Berkeley, 94720, USA.

* Corresponding author. Email: chengyitu1986@gmail.com (C. Tu), samir.suweis@unipd.it (S. Suweis).

[1] These authors contributed equally to this work.

# Abstract

The management of common-pool resources is a complex challenge due to the risk of overexploitation and the tragedy of the commons. A novel framework has been introduced to address this issue, focusing on the coevolutionary relationship between human behavior and common-pool resources within a human-environment system. However, the impact of the Allee effect on the coevolution and its resource sustainability is still unexplored. The Allee effect, a biological phenomenon characterized by a correlation between resource availability and growth rate, is a fundamental attribute of numerous natural resources. In this paper, we introduce two coevolutionary models of resource and strategy under replicator dynamics and knowledge feedback by applying the Allee effect to the common-pool resources within human-environment system. These models encapsulate various facets of resource dynamics and the players' behavior, such as resource growth function, the extraction rates, and the strategy update rules. We find that the Allee effect can induce bi-stability and critical transition, leading to either sustainable or unsustainable outcomes depending on the initial condition and parameter configuration. We demonstrate that knowledge feedback enhances the resilience and sustainability of the coevolving system, and these results advances the understanding of human-environment system and management of common-pool resources.

# Introduction

Common-pool resources (CPRs) are natural or human-made resources that are shared and exploited by multiple users, such as fisheries, forests, irrigation systems, or the atmosphere[1-7]. The management of CPRs poses a significant challenge for human societies, as they are subject to overexploitation and degradation due to the collective action problem and the tragedy of the commons[8,9]. In order to prevent the collapse of CPRs, it is necessary for users to cooperate and coordinate their actions to achieve a sustainable state where the CPRs are not exhausted. However, cooperation is often difficult to maintain in CPRs, as users face various incentives and constraints that may lead them to defect and free-ride on the efforts of others[1-4,10-14].

From a theoretical perspective, the intricate and dynamic interplay between human behaviors and resource can be modelled by a human-environment system (HES)[15-17]. To comprehend the evolution of the HES, it is essential to consider both the dynamic evolutionary of human behavior and resource[18-20]. Recently, a novel HES framework, based on two ordinary differential equations (ODEs) that capture the temporal evolution of human behaviors and resources, has been proposed[21-24]. The research assumes a logistic function without Allee effect[25,26] and demonstrates that users with shared goals exhibit a high level of self-organized cooperation, leading to long-term resource sustainability. In contrast, self-interested or individualistic behaviors result in resource depletion.

The Allee effect, a concept originally identified in the field of ecology, can be aptly applied to the context of CPRs. The Allee effect is a phenomenon that describes the relationship between population size and individual fitness[27-29]. It occurs when the population size is too low, making it harder for the individuals to find mates, reproduce, defend themselves from predators, or cooperate for mutual benefits[30,31]. The severity of the Allee effect can fluctuate, thereby determining a critical population threshold below which survival becomes unfeasible. This crucial threshold is termed the Allee parameter. When the population size falls below this critical threshold, the population experiences a reduced per capita growth rate, leading to a negative population growth rate and inevitable extinction[32,33]. Conversely, when the population size exceeds this critical threshold, the population growth rate is positive, leading to an increase in population size. When applying this concept to CPRs, the 'population' can be interpreted as the resource level of the CPRs[20,34,35]. In this analogy, the critical threshold could represent the minimum quantity of resources required to maintain sustainability. If the resources fall below this threshold, they are scarce, leading to resource exhaustion. Conversely, if the level of resource surpasses this threshold, their increased abundance leads to an increased per capita growth rate (meaning they become able to reproduce more effectively), allowing a sustainable utilization paradigm. Therefore, the Allee effect emerges as a more pragmatic determinant influencing the evolution and sustainability of CPRs. However, there is a gap in the literature on a novel and comprehensive HES framework that incorporates the Allee effect. The Allee parameter has significant implications for policy and practice as it can inform design and implementation of effective interventions to prevent or reverse resource degradation.

In this paper, we conduct a comprehensive analysis of the Allee effect, a biological phenomenon that exhibits a correlation between resource availability and growth rate, and its impact on the dynamics and long-term sustainability of CPRs. To capture the intricate interplay between resource availability and the behavior of the players, namely consumers, involved, we employ two distinct evolutionary models. These models allow us to examine the stability of the system and its sustainability under a variety of scenarios. Our finds reveal that the Allee effect can induce bi-stability and critical transitions within the system, leading to divergent outcomes, either sustainable or unsustainable, depending on the initial conditions and parameter configurations. Furthermore, we demonstrate that the knowledge feedback significantly enhances the resilience and sustainability of the co-evolving system. This insight propels forward our comprehension of HES and the management of CPRs by integrating the Allee effect. Overall, this paper contributes to the understanding of the complex and diverse dynamics of HES and significant implications for the management and policy of CPRs.

# Method

## Allee effect

In the context of natural resource management, the logistic function is often used to model the growth and utilization of renewable resources[25,26]. This function is mathematically expressed by the equation: $\frac{dR(t)}{dt} = TR(t)\left(1 - \frac{R(t)}{K}\right)$ where $R(t)$ is the size of the resource stock at time $t$, $T > 0$ is natural growth rate of the resource, $K$ is the carrying capacity. This function assumes that the resource growth rate is low when the resource is either scarce or abundant, and high when the resource is intermediate (see Fig. 1 a). However, this does not reflect the real-world scenario where the Allee effect may occur. The Allee effect, which is observed in some biological populations, implies that the resource growth rate may also depend on the resource state. We incorporated the Allee effect into the logistic function as follows: $\frac{dR(t)}{dt} = TR(t)\left(\frac{R(t)}{A} - 1\right)\left(1 - \frac{R(t)}{K}\right)$ where $0 < A < K$ is the Allee parameter. The Allee effect affects the resource growth rate depending on the value of resource $R$. Even when the Allee parameter $A$ is small, a critical value of $R$ exists below which the ratio $R/A - 1$ falls below zero, that is, for every instance where $R < A$. Upon the manifestation of the Allee effect, the Allee parameter $A$ governs the threshold, which invariably exists; in essence, strong and mild Allee effects are not qualitatively distinct, but only quantitatively. Specifically, if the Allee effect is mild, i.e., Allee parameter $A$ is small, the resource growth rate is low and the resource can persist at any positive level (see Fig. 1 b). If the Allee effect is strong, i.e., Allee parameter $A$ is large, the resource has a critical threshold below which the resource growth rate becomes negative so that the resource is doomed to extinction without any further aid (see Fig. 1 c). This observation underscores the intricate interplay between the Allee parameter and its impact on resource renewability.

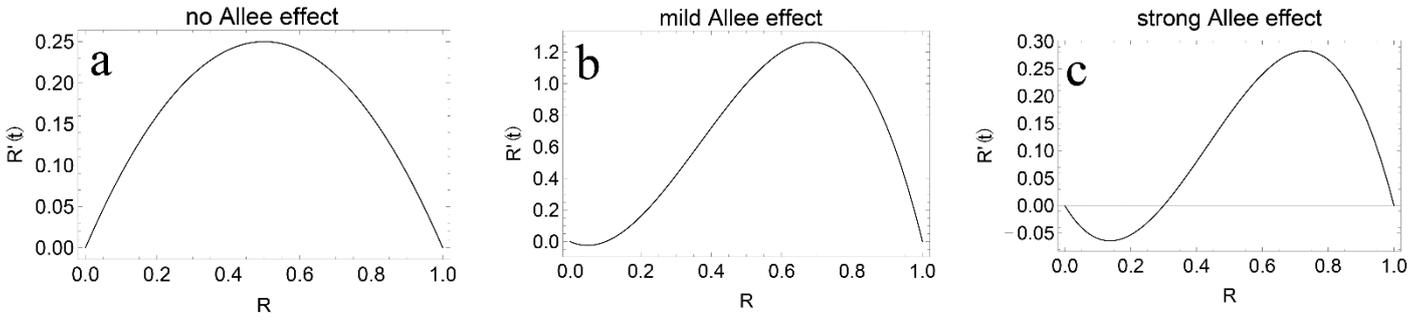

**Fig. 1 | An illustration of the Allee effect.** (a) The resource growth rate as a function of resource modelled by a logistic function; (b) The resource growth rate as a function of resource modelled by a logistic function with a mild Allee effect $A = 0.1$; (c) The resource growth rate as a function of resource modelled by a logistic function with a strong Allee effect $A = 0.3$. The other parameters are natural growth rate $T = 1$ and carrying capacity $K = 1$.

## Resource evolution

We model a dynamic resource pool whose volume $R(t)$ obeys a logistic function with Allee effect and is extracted by $N$ players. Each player has the option to employ one of two distinct strategies. The first is a cooperative strategy, characterized by the sustainable extraction of resources. If all members of the community adopt this strategy, the resource would maintain a non-zero state. Conversely, the second strategy is a defective one, which involves the extraction of resources at an unsustainable rate. If all members of the community adopt this strategy, the resource would inevitably be depleted. The fraction of cooperators in the system is denoted by $x = \frac{N_C}{N}$ where $N_C$ is number of cooperators (and the fraction of defectors is $1-x$). The extraction rates for cooperators and defectors are $e_C, e_D$, respectively, with $0 < Ne_C < T < Ne_D$. The total extraction rate by all players is $E = N_C e_C + N_D e_D = N(xe_C + (1-x)e_D)$. The ODE for the resource evolution is then given by

$$\begin{aligned}\frac{dR(t)}{dt} &= TR(t)\left(\frac{R(t)}{A}-1\right)\left(1-\frac{R(t)}{K}\right) - R(t)E = TR(t)\left(\frac{R(t)}{A}-1\right)\left(1-\frac{R(t)}{K}\right) - NR(t)\left(x(t)e_C + (1-x(t))e_D\right) \\ &= T\left[R(t)\left(\frac{R(t)}{A}-1\right)\left(1-\frac{R(t)}{K}\right) - R(t)\left(x(t)\hat{e}_C + (1-x(t))\hat{e}_D\right)\right]\end{aligned} \quad (1)$$

where $\hat{e}_C = \frac{Ne_C}{T}, \hat{e}_D = \frac{Ne_D}{T}$ are normalized extraction parameters. In the subsequent analysis, we establish $K$ as equal to 1 by normalizing the resource volume $R$ between 0 and 1 for simplicity. This decision is made in light of the fact that in the preceding equation, $K$ retains its general form. This adjustment allows for a more focused examination of the variables at play.

## Strategy evolution driven by replicator dynamics

In the scenario where players interact through a complete network, meaning each node is connected to all other nodes, the evolution of the probability $P^\tau(N_C)$, which represents the likelihood of having $N_C$ cooperators at time $\tau$, is dictated by following Master Equation[36,37]

$$P^{\tau+1}(N_C) - P^\tau(N_C) = P^\tau(N_C-1)T^{D \to C}(N_C-1 | R;\tau) + P^\tau(N_C+1)T^{C \to D}(N_C+1 | R;\tau) \\ - P^\tau(N_C)T^{C \to D}(N_C | R;\tau) - P^\tau(N_C)T^{D \to C}(N_C | R;\tau)$$

where $T^{D \to C}(N_C \pm 1 | R;\tau)$ is transition probability at time $\tau$ from $N_C$ to $N_C \pm 1$ given a resource level $R$.

Within the evolutionary game framework, we can derive a generalized Fokker-Planck equation for the probability density $\rho(x)$ of observing the cooperators fraction $x$. This derivation begins with the Master Equation, and involves the substitutions $x = \frac{N_C}{N}, t = \frac{\tau}{N}$ and $\rho(x;t) = NP^\tau(N_C)$, leading to the following equation:

$$\rho(x;t+N^{-1}) - \rho(x;t) = \rho(x-N^{-1};t)T^{D \to C}(x-N^{-1} | R;t) + \rho(x+N^{-1};t)T^{C \to D}(x+N^{-1} | R;t) \\ - \rho(x;t)T^{C \to D}(x | R;t) - \rho(x;t)T^{D \to C}(x | R;t)$$

In the case where $N$ is significantly larger than 1, we can employ a Taylor series expansion around $x(t)$ to approximate both the probability densities and the transition probabilities. By disregarding terms of higher order in $N^{-1}$, we can derive the Fokker-Planck equation as follows:

$$\frac{d}{dt}\rho(x;t) = -\frac{d}{dx}[a(x | R;t)\rho(x;t)] + \frac{1}{2}\frac{d^2}{dx^2}[b^2(x | R;t)\rho(x;t)]$$

where $a(x | R;t) = T^{D \to C}(x | R;t) - T^{C \to D}(x | R;t)$ and $b(x | R;t) = \sqrt{\frac{1}{N}\left(T^{D \to C}(x | R;t) + T^{C \to D}(x | R;t)\right)}$.

Given that the time steps in this model are independent, resulting in uncorrelated noise over time, we can utilize Ito calculus to derive the corresponding Langevin equation as follows:

$$\frac{dx(t)}{dt} = a(x | R;t) + b(x | R;t)\zeta$$

where $\zeta$ is an uncorrelated Gaussian white noise. In the limit as $N \to \infty$, the diffusion term $b(x | R;t)$ tends to zero with a rate of $1/\sqrt{N}$, resulting in a deterministic equation. Consequently, the evolutionary dynamics of the players' strategies can be expressed as follows:

$$\frac{dx(t)}{dt} = T^{D \to C} - T^{C \to D} \quad (2)$$

where $T^{D \to C}, T^{C \to D}$ is transition probability for Master Equation[38,39].

We start from the simplest case of the replicator dynamics[40,41], which models the players' strategy as follows: 1) a player

$i$ is randomly chosen; 2) a neighbor of $i$ is randomly chosen, $j$; 3) the strategy of player $i$ switches to the strategy of player $j$ with probability $p = \frac{1}{2} + \frac{w}{2}\frac{U_j - U_i}{\Delta U_{max}}$ where $U_i, U_j$ are the payoffs of players $i$ and $j$ which are directly proportional to the quantity of resources extracted by each player, expressed as $U_X = Re_X$. In this equation, $e_X$ signifies the extraction rate, while $X$ denotes the player's chosen strategy, either cooperation ($X = C$) and defection ($X = D$). This mathematical representation elucidates the dynamics of strategic decision-making in resource extraction scenarios. $\Delta U_{max}$ is the maximum possible payoff difference that ensures $0 \leq p \leq 1$ ($\Delta U_{max} = \max |U_D - U_C| = e_D - e_C$ when $R = 1$) and $0 < w \leq 1$ is the greed parameter[21]. In this setting, the probabilities to switch strategy from defection to cooperation, and vice versa are $p^{D \to C} = \frac{1}{2} + \frac{w}{2}\frac{Re_C - Re_D}{e_D - e_C} = \frac{1}{2} - \frac{w}{2}R$ and $p^{C \to D} = \frac{1}{2} + \frac{w}{2}\frac{Re_D - Re_C}{e_D - e_C} = \frac{1}{2} + \frac{w}{2}R$, respectively, and the transition probabilities are $T^{D \to C} = x(1-x)p^{D \to C}$ and $T^{C \to D} = x(1-x)p^{C \to D}$. The evolutionary dynamics of players' strategies under the replicator dynamics is

$$\frac{dx(t)}{dt} = -w(t)R(t)(1-x(t))x(t) \quad (3)$$

The greed parameter reflects how much a player prefers to extract more resource by defection than the sustainable level by cooperation. When $w > 0$, the fraction of cooperators $x$ decreases and the fraction of defectors $1-x$ increases. Moreover, the larger the absolute value of $w$, the faster the evolution of the players' strategy.

## Resource and strategy coevolution under replicator dynamics

In order to capture the comprehensive dynamics of the HES, we put together the resource evolution given by Eq. (1) and the strategy evolution under replicator dynamics given by Eq. (3). This amalgamation yields the subsequent coevolutionary dynamics of resource and strategy, as given by

$$\begin{cases} \frac{dR(t)}{dt} = T\left(R(t)\left(\frac{R(t)}{A} - 1\right)\left(1 - \frac{R(t)}{K}\right) - R(t)\left(x(t)\hat{e}_C + (1-x(t))\hat{e}_D\right)\right) \\ \frac{dx(t)}{dt} = -wR(t)x(t)(1-x(t)) \end{cases} \quad (4)$$

where $0 \leq R(t) \leq 1$ is resource volume, $0 \leq x(t) \leq 1$ is cooperators fraction, $0 < w \leq 1$ is greed parameter, $\hat{e}_C = \frac{Ne_C}{T}, \hat{e}_D = \frac{Ne_D}{T}$ are normalized extraction rates of cooperators and defectors where $0 < \hat{e}_C < 1 < \hat{e}_D$ and $K$ can set to 1 by normalizing $R$ for simplicity.

The macroscopic ODE captures the population-level dynamics of the fraction of cooperators and resource, while the microscopic update reflects the switching of the players' strategies at the individual level and the subsequent resource update. Our macroscopic ODE is supplemented with a microscopic update of the resource and strategy coevolution under replicator dynamics. The microscopic update after the initialization consist of the following steps: 1) a player $i$ is randomly chosen; 2) a neighbor of $i$ is randomly chosen, $j$; 3) the strategy of player $i$ switches to the strategy of player $j$ with probability $p = \frac{1}{2} + \frac{w}{2}\frac{U_j - U_i}{\Delta U_{max}}$; 4) update $x[k] = \frac{N_C}{N}$ where $k$ is the discrete time; then the resource dynamics can be expressed as

$$R[k] = R[k-1] + \frac{T}{N}\left( R[k-1]\left(\frac{R[k-1]}{A} - 1\right)\left(1 - \frac{R[k-1]}{K}\right) - R[k-1]\left(x[k-1]\hat{e}_C + (1-x[k-1])\hat{e}_D\right) \right),$$

which is simply the discretization of Eq. (4). A comparative analysis of the microscopic update and macroscopic ODE under various parameter configurations reveals a consistency in the results (see Fig. 2).

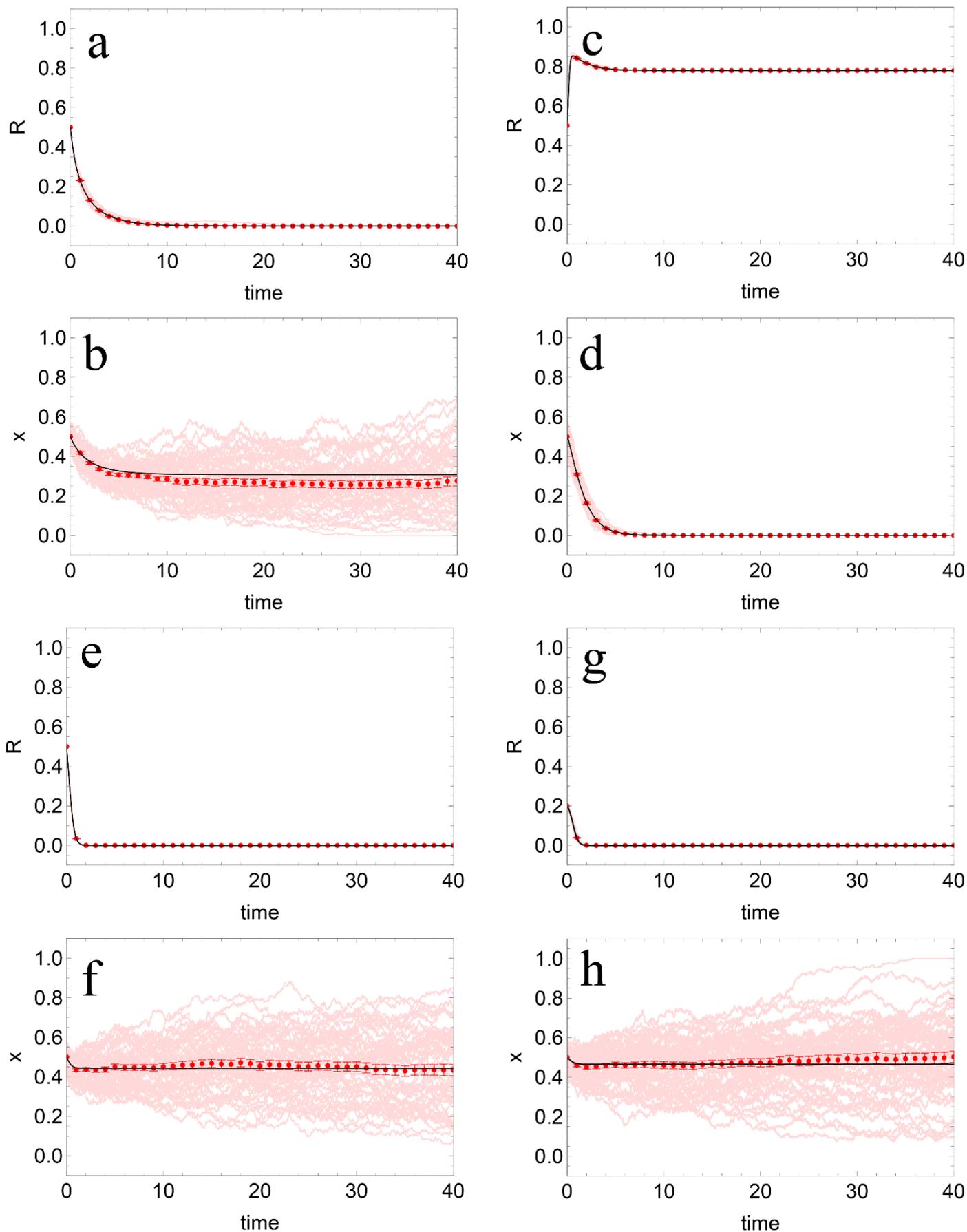

**Fig. 2 | Comparison between microscopic update and macroscopic ODE for resource and strategy coevolution under replicator dynamics.** (a-b) The evolution of the resource $R$ and cooperator fraction $x$ without Allee effect with initial condition $R_0 = 0.5, x_0 = 0.5$; (c-d) The evolution of the resource $R$ and cooperator fraction $x$ with a mild Allee effect

$A = 0.1$ with initial condition $R_0 = 0.5, x_0 = 0.5$; (e-f) The evolution of the resource $R$ and cooperator fraction $x$ with a strong Allee effect $A = 0.3$ with initial condition $R_0 = 0.5, x_0 = 0.5$; (g-h) The evolution of the resource $R$ and cooperator fraction $x$ with a mild Allee effect $A = 0.1$ with initial condition $R_0 = 0.2, x_0 = 0.5$. The other parameters are population size $N = 200$, greed parameter $w = 1$, growth rate $T = 2$, normalized extraction parameter $\hat{e}_C = 0.5, \hat{e}_D = 1.5$. We performed $n = 50$ independent numerical simulations for each case, shown as light red lines. The red dots and error bars indicate the mean and standard error of the simulations at each discrete time step. The black line represents the analytical solution of the macroscopic ODE.

## Sustainability for coevolution under replicator dynamics

We perform a theoretical analysis of the sustainability of resource and strategy coevolution under replicator dynamics given by Eq. (4), using the steady-state solutions that describe the system's asymptotic behavior. We first examine the stationary solutions of the coupled equations $\{R^*, x^*\}$, which are given by $\mathbf{s}_0 = \{R = 0, \forall x\}$, $\mathbf{s}_{00} = \{R = 0, x = 0\}$,

$\mathbf{s}_{01} = \{R = 0, x = 1\}$ , $\mathbf{s}_{C^{\mp}} = \{R = \frac{1}{2}\left(1 + A \mp \sqrt{(1-A)^2 - 4A\hat{e}_C}\right), x = 1\}$ existing for $\hat{e}_C \leq \frac{(1-A)^2}{4A}$ and

$\mathbf{s}_{D^{\mp}} = \{R = \frac{1}{2}\left(1 + A \mp \sqrt{(1-A)^2 - 4A\hat{e}_D}\right), x = 0\}$ existing for $\hat{e}_D \leq \frac{(1-A)^2}{4A}$.

We then analyze the stability of these solutions for the parameter configuration $T, A, \hat{e}_C, \hat{e}_D, w$ and initial condition $R_0, x_0$. The eigenvalues of the Jacobian matrix at these solutions are: $\boldsymbol{\lambda}_0 = \{0, -T\left(1 + \hat{e}_D(1-x^*) + \hat{e}_C x^*\right)\}$,

$\boldsymbol{\lambda}_{00} = \{-T(1+\hat{e}_D), 0\}$ , $\boldsymbol{\lambda}_{01} = \{-T(1+\hat{e}_C), 0\}$ ,

$\boldsymbol{\lambda}_{C^{\mp}} = \{\frac{T}{2A}\left(-((1-A)^2 - 4A\hat{e}_C) \pm (1+A)\sqrt{(1-A)^2 - 4A\hat{e}_C}\right), \frac{w}{2}\left(1 + A \mp \sqrt{(1-A)^2 - 4A\hat{e}_C}\right)\}$ and

$\boldsymbol{\lambda}_{D^{\mp}} = \{\frac{T}{2A}\left(-((1-A)^2 - 4A\hat{e}_D) \pm (1+A)\sqrt{(1-A)^2 - 4A\hat{e}_D}\right), -\frac{w}{2}\left(1 + A \mp \sqrt{(1-A)^2 - 4A\hat{e}_D}\right)\}$. Since

$1 + \hat{e}_D(1-x^*) + \hat{e}_C x^* > 0$ for any $x^*$, $\mathbf{s}_0$ is always a neutrally stable solution. The solutions $\mathbf{s}_{00}$ and $\mathbf{s}_{01}$ exist if and only if $x_0 = 0$ and $x_0 = 1$. Because $1 + A \mp \sqrt{(1-A)^2 - 4A\hat{e}_X} > 0$, $(\boldsymbol{\lambda}_{D^{\mp}})_2 < 0$ and $(\boldsymbol{\lambda}_{C^{\mp}})_2 > 0$. Therefore, $\mathbf{s}_{C^{\mp}}$ are unstable. When $\hat{e}_D > \frac{(1-A)^2}{4A}$, $\mathbf{s}_{D^{\mp}}$ does not exist and $\mathbf{s}_0$ is the only stable solution. When $\hat{e}_D < \frac{(1-A)^2}{4A}$, $(\boldsymbol{\lambda}_{D^-})_1 > 0$ due to $-((1-A)^2 - 4A\hat{e}_C) + (1+A)\sqrt{(1-A)^2 - 4A\hat{e}_C} > 0$ and $(\boldsymbol{\lambda}_{D^+})_1 < 0$ due to

$-\left((1-A)^2 - 4A\hat{e}_C\right) - (1+A)\sqrt{(1-A)^2 - 4A\hat{e}_C} < 0$. Therefore, $\mathbf{s}_{D^-}$ is unstable and $\mathbf{s}_{D^+}$ is stable. The system exhibits bi-stability between $\mathbf{s}_0$ and $\mathbf{s}_{D^+}$ depending on the initial condition $\{R_0, x_0\}$ (see Fig. 2 c-d and g-h). It is particularly noteworthy to highlight that even if all players are defectors finally, i.e., $x^* = 0$, the sustainability of the resource is not compromised, i.e., $R^* > 0$ (see Fig. 2 c-d).

We approximate the boundaries of the basins of attraction by examining the special case where $x_0 = 0$ and $x_0 = 1$ and then interpolating between these extremes. For $x = 1$, the resource dynamics is given by $\frac{dR(t)}{dt} = T\left(R(t)\left(\frac{R(t)}{A} - 1\right)(1 - R(t)) - R(t)\hat{e}_C\right)$. Its stationary solutions $R^*$ are: $(\mathbf{s}_{00})_R = 0$, $(\mathbf{s}_{C^\mp})_R = \frac{1}{2}\left(1 + A \mp \sqrt{(1-A)^2 - 4A\hat{e}_C}\right)$ and the corresponding eigenvalues evaluated from the Jacobian matrix are: $(\lambda_{00})_R < 0$, $(\lambda_{C^-})_R > 0$ and $(\lambda_{C^+})_R < 0$. Therefore, $\mathbf{s}_{C^-}$ is a critical point to distinguish the attractor of $\mathbf{s}_{00}$ and $\mathbf{s}_{C^+}$. For $x = 0$, the resource dynamics is given by $\frac{dR(t)}{dt} = T\left(R(t)\left(\frac{R(t)}{A} - 1\right)(1 - R(t)) - R(t)\hat{e}_D\right)$. Its stationary solutions $R^*$ are: $(\mathbf{s}_{00})_R = 0$, $(\mathbf{s}_{D^\mp})_R = \frac{1}{2}\left(1 + A \mp \sqrt{(1-A)^2 - 4A\hat{e}_D}\right)$ and the corresponding eigenvalues evaluated from its Jacobian matrix are: $(\lambda_{01})_R < 0$, $(\lambda_{D^-})_R > 0$ and $(\lambda_{D^+})_R < 0$. Therefore, $\mathbf{s}_{D^-}$ is another critical point. The initial condition $\{R_0, x_0\}$ determines the global stability of the system. If it lies to the left of the line composed by $\mathbf{s}_{C^-}$ and $\mathbf{s}_{D^-}$, i.e., $x_0 = \frac{R_0 - (\mathbf{s}_{D^-})_R}{(\mathbf{s}_{C^-})_R - (\mathbf{s}_{D^-})_R}$, then the system converges to $\mathbf{s}_0$; otherwise, it converges to $\mathbf{s}_{D^+}$.

## Resource and strategy coevolution under knowledge feedback

We explore a scenario where players exhibit rational behavior, choosing to defect only when the advantages of defection surpass the Allee parameter $A$. This threshold signifies the minimal level of cooperation necessary for the sustainability of the resource. The role of knowledge feedback is emphasized, where the player's decision-making process is influenced by both the resource state and the Allee effect. We abandon the replicator dynamics and its associated greed parameter $w$, and introduce resource and Allee effect from resource evolution to strategy evolution. Specifically, we define a new probability of switching from cooperation to defection $p^{C \to D} = p = \theta[R - A]\left(\frac{R - A}{K - A}\right)$ and probability of switching from defection to cooperation $p^{D \to C} = 1 - p$ where $\theta[x]$ represents the unit step function, which is 0 for

$x < 0$ and 1 for $x \geq 0$. The probabilities of strategy switching depend on the value of the resource $R$ and Allee parameter $A$. When the resource $R$ is lower than Allee parameter $A$, the probability of switching from cooperation to defection $p^{C \to D}$ is 0 while the probability of switching from defection to cooperation $p^{D \to C}$ is 1. This means that all players will adopt cooperation as their strategy for resource sustainability, until the resource $R$ increases and finally touches $A$. When the resource $R$ exceeds Allee parameter $A$, the probability of switching from cooperation to defection $p^{C \to D}$ is $\frac{R-A}{K-A}$ while the probability of switching from defection to cooperation $p^{D \to C}$ is its complement. As resource $R$ approaches Allee parameter $A$, the probability of switching from cooperation to defection $p^{C \to D}$ decreases and its complement $p^{D \to C}$ increases. The evolutionary dynamics of players' strategies under knowledge feedback is given by

$$\frac{dx(t)}{dt} = T^{D \to C} - T^{C \to D} = 1 - x - p = 1 - x - \theta[R-A]\left(\frac{R-A}{K-A}\right) \quad (5)$$

In order to capture the comprehensive dynamics of the HES under knowledge feedback, we integrate the resource evolution given by Eq. (1) and the strategy evolution under knowledge feedback given by Eq. (5). This amalgamation yields the subsequent coevolutionary dynamics of resource and strategy, as given by

$$\begin{cases} \frac{dR(t)}{dt} = T\left(R(t)\left(\frac{R(t)}{A}-1\right)\left(1-\frac{R(t)}{K}\right) - R(t)\left(x(t)\hat{e}_C + (1-x(t))\hat{e}_D\right)\right) \\ \frac{dx(t)}{dt} = 1 - x - \theta[R-A]\left(\frac{R-A}{K-A}\right) \end{cases} \quad (6)$$

where $K$ can be set to 1 by normalizing $R$ for simplicity.

The macroscopic ODE captures the population-level dynamics of the fraction of cooperators and resource, while the microscopic update reflects the individual-level strategy switching of the players and the subsequent resource update. Our macroscopic ODE is supplemented with a microscopic update of the resource and strategy coevolution under knowledge feedback. The microscopic update after the initialization consist of the following steps: 1) a player $i$ is randomly chosen; 2) if player $i$ is cooperator, it switches to defector with probability $p^{C \to D}$; if $i$ is defector, it switches to cooperator with probability $p^{D \to C}$; 3) update $x[k] = \frac{N_C}{N}$ where $k$ is the discrete time; then the resource dynamics can be expressed as $R[k] = R[k-1] + \frac{T}{N}\left(R[k-1]\left(\frac{R[k-1]}{A}-1\right)\left(1-\frac{R[k-1]}{K}\right) - R[k-1]\left(x[k-1]\hat{e}_C + (1-x[k-1])\hat{e}_D\right)\right)$, which is simply the discretization of Eq. (4). A comparative analysis of the microscopic update and macroscopic ODE under various parameter configurations reveals a consistency in the results (see Fig. 3).

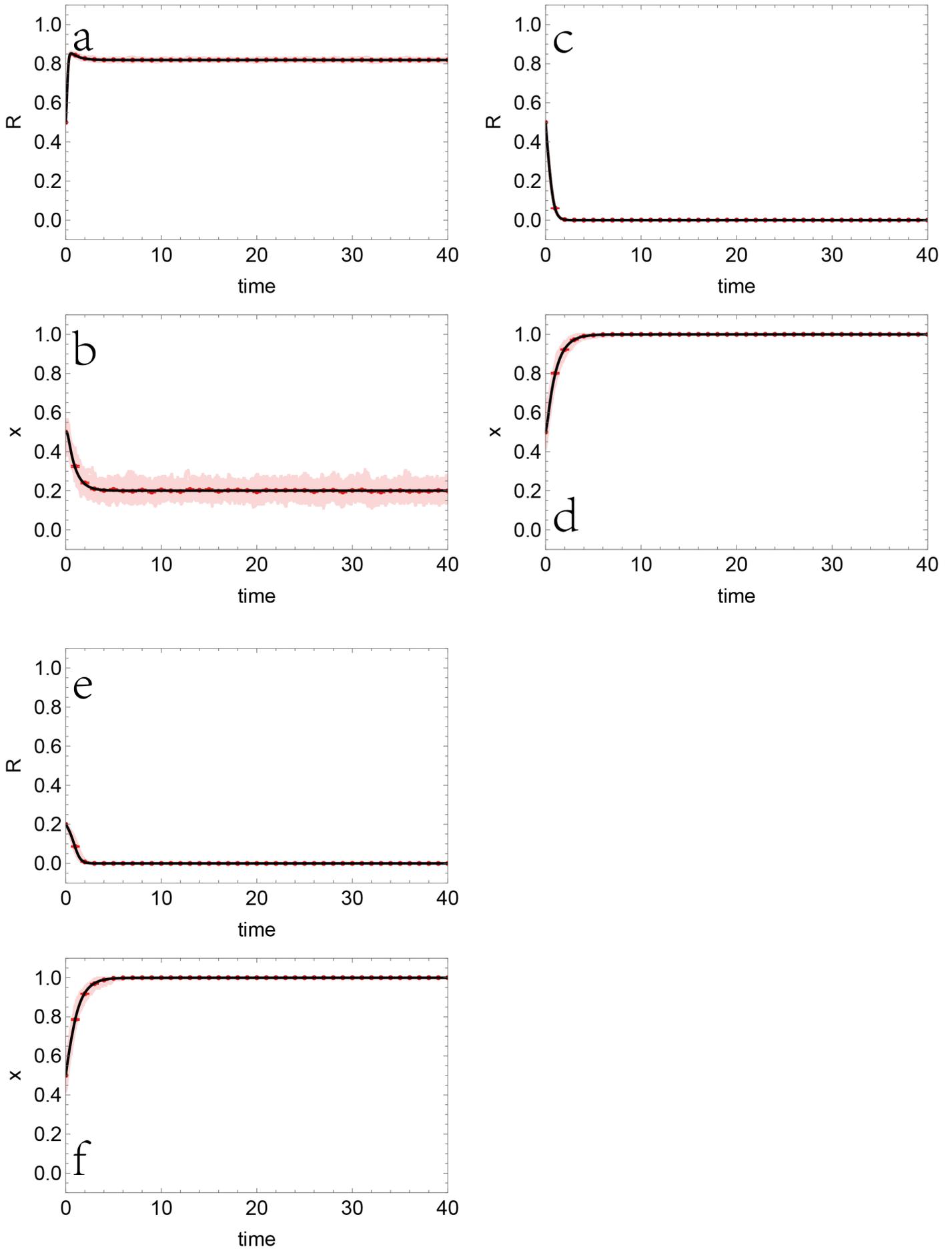

**Fig. 3 | Comparison between microscopic update and macroscopic ODE for resource and strategy coevolution under knowledge feedback.** (a-b) The evolution of the resource $R$ and cooperator fraction $x$ with a mild Allee effect $A = 0.1$

with initial condition $R_0 = 0.5, x_0 = 0.5$; (c-d) The evolution of the resource $R$ and cooperator fraction $x$ with a strong Allee effect $A = 0.3$ with initial condition $R_0 = 0.5, x_0 = 0.5$; (e-f) The evolution of the resource $R$ and cooperator fraction $x$ with a mild Allee effect $A = 0.1$ with initial condition $R_0 = 0.2, x_0 = 0.5$. The other parameters are population size $N = 200$, growth rate $T = 2$, normalized extraction parameter $\hat{e}_C = 0.5, \hat{e}_D = 1.5$. We performed $n = 50$ independent numerical simulations for each case, shown as light red lines. The red dots and error bars indicate the mean and standard error of the simulations at each discrete time step. The black line represents the analytical solution of the macroscopic ODE.

## Sustainability for coevolution under knowledge feedback

We perform a theoretical analysis of the sustainability of resource and strategy coevolution under knowledge feedback given by Eq. (6), using the steady-state solutions that describe the system's asymptotic behavior. We first examine the stationary solutions of the coupled equations $\{R^*, x^*\}$, which are given by: $\mathbf{s}_0 = \{R = 0, x = 1\}$,

$$\mathbf{s}_- = \{R = \frac{-1 + A(A - \hat{e}_C + \hat{e}_D) - \sqrt{(A^2 - A(\hat{e}_C + 2) + 1)^2 + A^2\hat{e}_D^2 - 2A\hat{e}_D(A(A + \hat{e}_C - 2) + 1)}}{2(A-1)},$$

$$x = \frac{1 + A(A - \hat{e}_C + \hat{e}_D - 2) - \sqrt{(A^2 - A(\hat{e}_C + 2) + 1)^2 + A^2\hat{e}_D^2 - 2A\hat{e}_D(A(A + \hat{e}_C - 2) + 1)}}{2(A-1)^2}\}$$

existing for

$$\left(0 < A < 3 - 2\sqrt{2} \land 2\sqrt{\frac{(A-1)^2 \hat{e}_C}{A}} - A - \frac{1}{A} - \hat{e}_C + \hat{e}_D + 2 < 0\right) \lor$$

$$\left(3 - 2\sqrt{2} < A < \frac{1}{2}(3 - \sqrt{5}) \land 2\sqrt{\frac{(A-1)^2 \hat{e}_C}{A}} - A - \frac{1}{A} - \hat{e}_C + \hat{e}_D + 2 < 0 \land -A + 2\sqrt{\frac{(A-1)^2}{A}} - \frac{1}{A} + \hat{e}_C + 1 < 0\right)$$

and

$$\mathbf{s}_+ = \{R = \frac{-1 + A(A - \hat{e}_C + \hat{e}_D) + \sqrt{(A^2 - A(\hat{e}_C + 2) + 1)^2 + A^2\hat{e}_D^2 - 2A\hat{e}_D(A(A + \hat{e}_C - 2) + 1)}}{2(A-1)},$$

$$x = \frac{1 + A(A - \hat{e}_C + \hat{e}_D - 2) + \sqrt{(A^2 - A(\hat{e}_C + 2) + 1)^2 + A^2\hat{e}_D^2 - 2A\hat{e}_D(A(A + \hat{e}_C - 2) + 1)}}{2(A-1)^2}\}$$

existing for the same region as $\mathbf{s}_-$. Using the method of $\mathrm{Det} > 0 \land \mathrm{Tr} < 0$, we determine that $\mathbf{s}_0$ is stable, $\mathbf{s}_-$ is unstable and $\mathbf{s}_+$ is stable.

# Results

## Allee effect for resource and strategy coevolution

We conduct numerical simulations to examine the coevolution of resource and strategy at both the macroscopic ODE and microscopic update level, using the same population size $N$, greed parameter $w$, growth rate $T$, normalized extraction parameter $\hat{e}_C, \hat{e}_D$ and initial condition $\{R_0, x_0\}$. In our previous work[21], we find that, in the absence of the Allee effect, the resource $R$ approximates to 0 and the cooperators fraction $x$ decreases over time (see Fig. 2 a-b). Conversely, when the Allee effect is mild (e.g., Allee parameter $A = 0.1$), the resource $R$ could persist even when the cooperator fraction $x$ is extremely low, for both the replicator dynamics and the knowledge feedback scenarios (see Fig. 2 c-d and Fig. 3 a-b). However, when the Allee effect is strong (e.g., Allee parameter $A = 0.3$), the resource $R$ is endangered even when the cooperator fraction $x$ is relatively high (see Fig. 2 e-f and Fig. 3 c-d). These results indicate that the Allee effect plays a crucial role in shaping the coevolutionary dynamics and outcomes of resource sustainability, and that its strength can have different effects depending on the feedback mechanism.

Additionally, our study indicates that when a mild Allee effect is present and the initial condition for resource $R$ is sufficiently high, sustainability can be realized (see Fig. 2 c and Fig. 3 a), a conclusion that stands in contrast to the logistic model (see Fig. 2 a). In contrast, when the Allee effect maintains the same intensity but is associated with a lower initial condition for resource $R$, resource depletion occurs (see Fig. 2 g and Fig. 3 e). Furthermore, if the Allee effect is pronounced, the resource $R$ is inevitably exhausted (see Fig. 2 e and Fig. 3 c). These findings highlight the intricate relationship between the Allee effect and initial conditions in the context of resource management.

## Critical $\hat{e}_D$ and $A$ for coevolution under replicator dynamics

The stability of the solutions of the resource and strategy coevolution under replicator dynamics depends on both the parameter configuration and the initial condition of the system. A region of bi-stability emerges when $\hat{e}_D < \frac{(1-A)^2}{4A}$. For this region (see Fig. 4 a), the system exhibits two distinct regimes of stability, separated by a critical line $x_0 = \frac{R_0 - (\mathbf{s}_{D^-})_R}{(\mathbf{s}_{C^-})_R - (\mathbf{s}_{D^-})_R}$ in the phase space (see Fig. 4 b). Below this line, the initial condition $\{R_0, x_0\}$ leads the system to converge to the unsustainable solution $\mathbf{s}_0$. Above this line, the system converges to the sustainable solution $\mathbf{s}_{D^+}$. Otherwise (see Fig. 4 c and e), the system has no sustainable solution and always converges to the unsustainable solution

$s_0$ (see Fig. 4 d and f). This indicates that the system is highly sensitive to the relationship between $A$ and $\hat{e}_D$. We find that the Allee effect can enhance or hinder the resource sustainability, depending on the parameter values and the initial conditions.

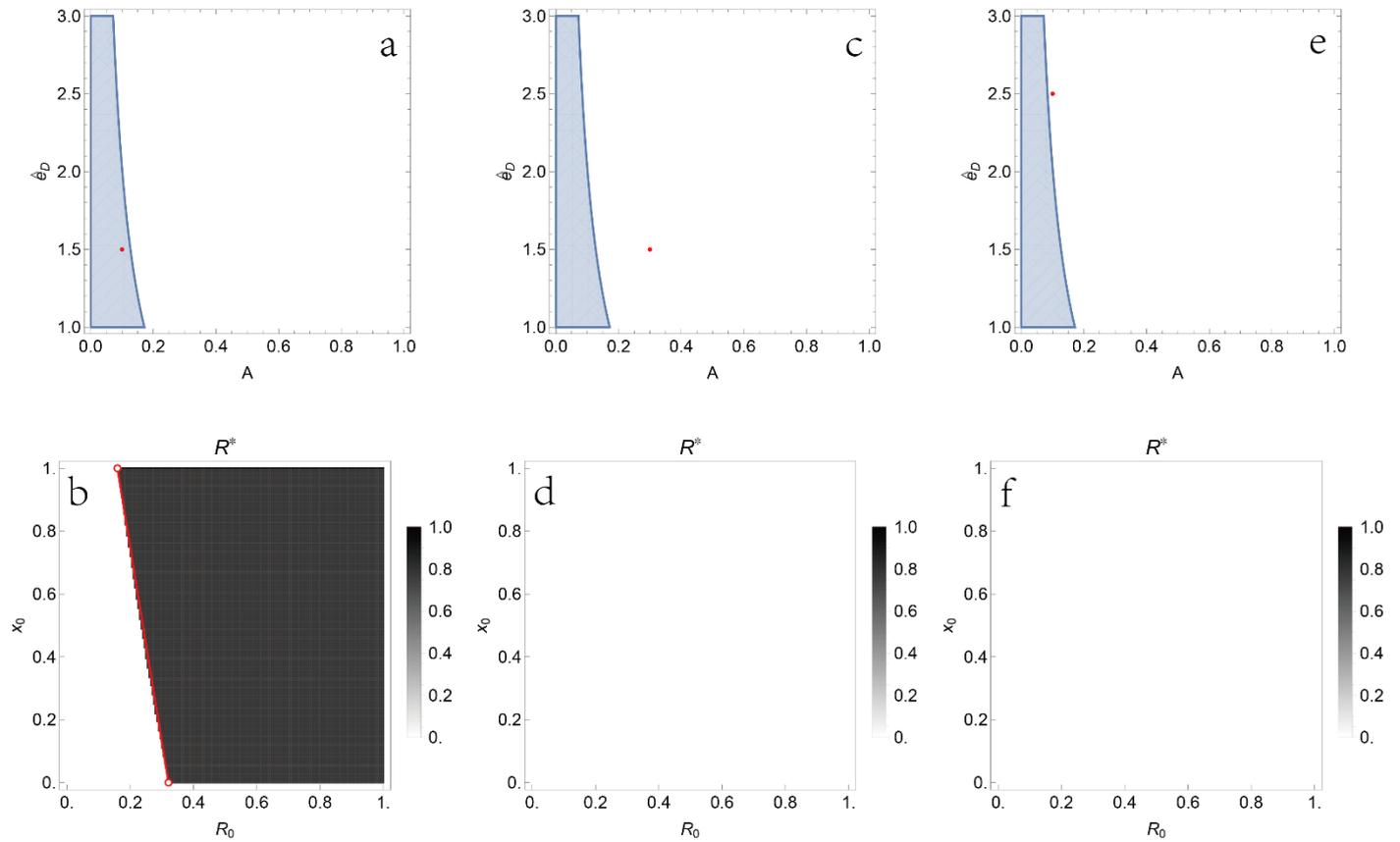

**Fig. 4 | Steady-state behavior of resource and strategy coevolution under replicator dynamics.** (a) The bi-stability condition is satisfied by the parameter values $A = 0.1$ and $\hat{e}_C = 0.5, \hat{e}_D = 1.5$, as shown by the red point in the blue region; (b) Density plots of the steady-state solutions for $R^*$, given the same values of $A$ and $\hat{e}_D$ as in panel a. The red line denotes the critical line, which separates the initial conditions that lead to different outcomes; (c) The bi-stability condition is not satisfied by the parameter values $A = 0.3$ and $\hat{e}_C = 0.5, \hat{e}_D = 1.5$; (d) Density plots of the steady-state solutions for $R^*$, given the same values of $A$ and $\hat{e}_D$ as in panel c; (e) The bi-stability condition is not satisfied by the parameter values $A = 0.1$ and $\hat{e}_C = 0.5, \hat{e}_D = 2.5$; (f) Density plots of the steady-state solutions for $R^*$, given the same values of $A$ and $\hat{e}_D$ as in panel e.

## Critical $\hat{e}_D$ and $A$ for coevolution under knowledge feedback

The stability of the of solutions of the resource and strategy coevolution under knowledge feedback depends on both the parameter configuration and the initial condition of the system. A region of bi-stability emerges when

$$\left(0 < A < 3 - 2\sqrt{2} \wedge 2\sqrt{\frac{(A-1)^2 \hat{e}_C}{A}} - A - \frac{1}{A} - \hat{e}_C + \hat{e}_D + 2 < 0\right) \vee$$
$$\left(3 - 2\sqrt{2} < A < \frac{1}{2}(3 - \sqrt{5}) \wedge 2\sqrt{\frac{(A-1)^2 \hat{e}_C}{A}} - A - \frac{1}{A} - \hat{e}_C + \hat{e}_D + 2 < 0 \wedge -A + 2\sqrt{\frac{(A-1)^2}{A}} - \frac{1}{A} + \hat{e}_C + 1 < 0\right).$$

For this region (see Fig. 5 a and e), the system exhibits bi-stability depending on the initial condition $\{R_0, x_0\}$ (see Fig. 5 b and f). Otherwise (see Fig. 5 c), the system has no sustainable solution and always converges to the unsustainable solution $\mathbf{s}_0$ (see Fig. 5 d).

We investigate the effects of knowledge feedback on the resource and strategy coevolution, and compare them with those of replicator dynamics. We delineate a bi-stability region of coevolution under replicator dynamics for any given value of $\hat{e}_C$ (see Fig. 6 a), and under knowledge feedback for distinct values of $\hat{e}_C = 0.25, 0.5, 0.75$ (see Fig. 6 b-d). Our findings reveal that, regardless of the value of $\hat{e}_C$, the sustainable region of knowledge feedback is consistently larger than that of replicator dynamics (see Fig. 6 a-d). Furthermore, when the system is in a bi-stable state for the same parameters for both scenarios, the basin of attraction of the sustainable equilibrium under knowledge feedback is consistently larger than that under replicator dynamics (see Fig. 6 e-h). These observations imply that knowledge feedback bolsters the sustainability of the coevolving system.

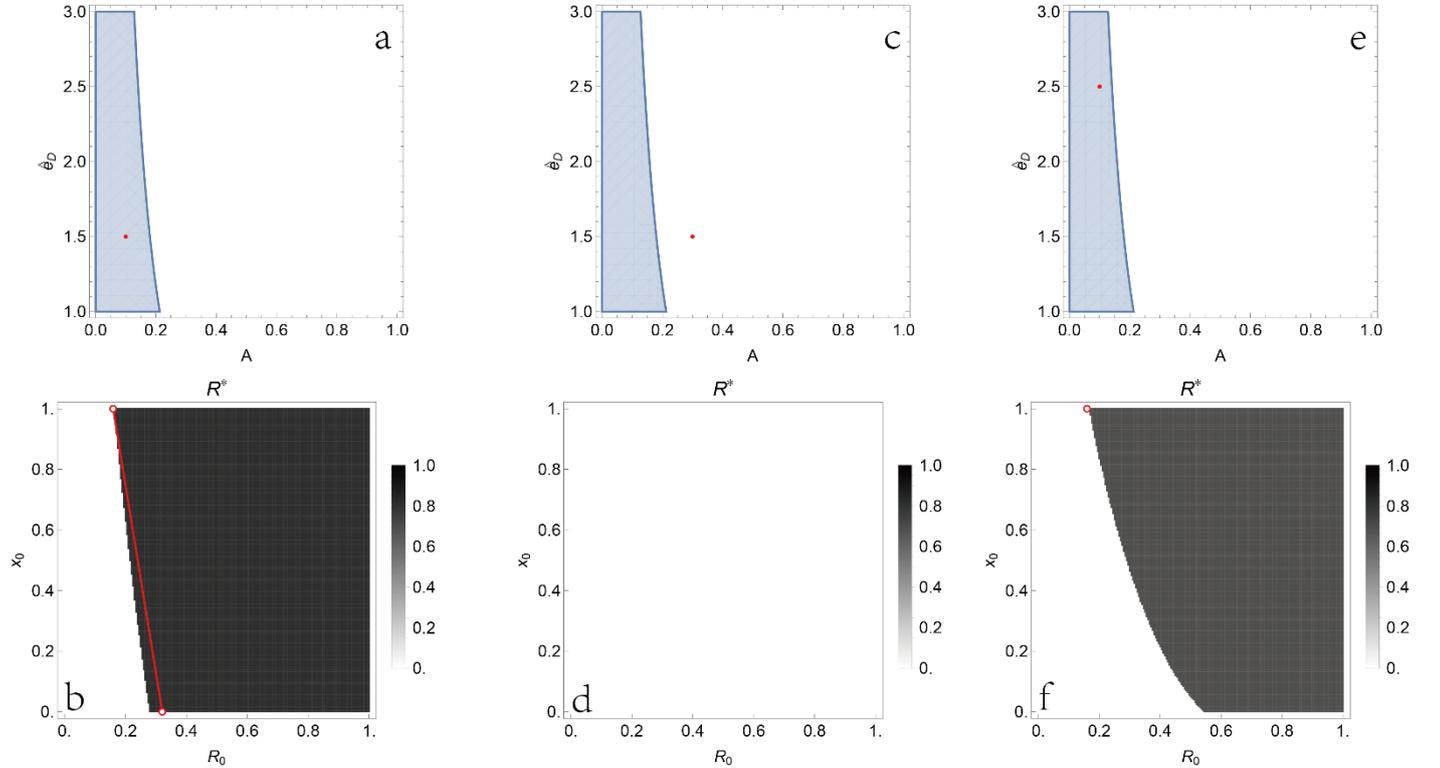

**Fig. 5 | Steady-state behavior of resource and strategy coevolution under knowledge feedback.** (a) The bi-stability condition is satisfied by the parameter values $A = 0.1$ and $\hat{e}_C = 0.5, \hat{e}_D = 1.5$, as shown by the red point in the blue region; (b) Density plots of the steady-state solutions for $R^*$, given the same values of $A$ and $\hat{e}_D$ as in panel a. The red

line denotes the critical line obtained in resource and strategy coevolution with replicator dynamics, which separates the initial conditions that lead to different outcomes; (c) The bi-stability condition is not satisfied by the parameter values $A = 0.3$ and $\hat{e}_C = 0.5, \hat{e}_D = 1.5$; (d) Density plots of the steady-state solutions for $R^*$, given the same values of $A$ and $\hat{e}_D$ as in panel c; (e) The bi-stability condition is satisfied by the parameter values $A = 0.1$ and $\hat{e}_C = 0.5, \hat{e}_D = 2.5$; (f) Density plots of the steady-state solutions for $R^*$, given the same values of $A$ and $\hat{e}_D$ as in panel e.

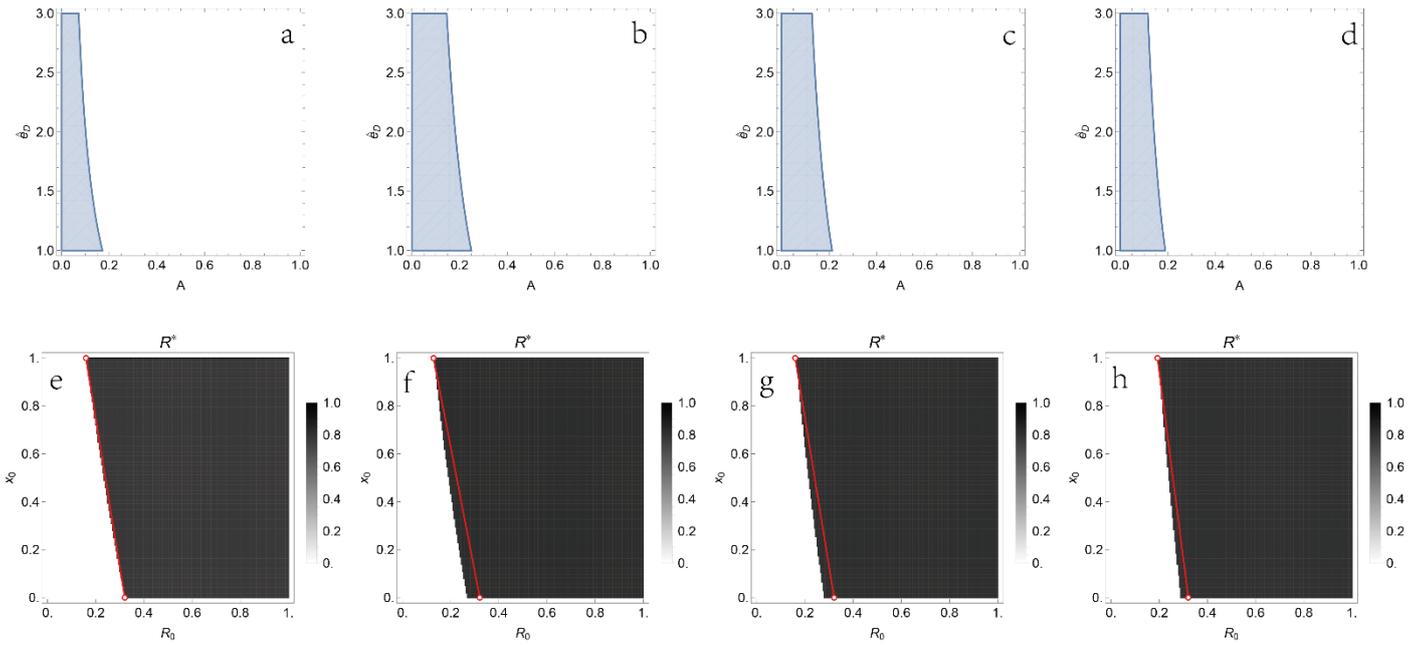

**Fig. 6 | Comparison between bi-stability of coevolution under replicator dynamics and knowledge feedback.** (a-d) The bi-stability region of coevolution under replicator dynamics as well as coevolution under knowledge feedback for different value of $\hat{e}_C = 0.25, 0.5, 0.75$. The other parameters are fixed at $A = 0.1$ and $\hat{e}_D = 1.5$; (e-h) Density plots of the steady-state solutions for $R^*$, given the same values of $A$ and $\hat{e}_D$ as in panel a-d. The red line denotes the critical line obtained in coevolution with replicator dynamics, which separates the initial conditions that lead to different outcomes.

## Universal critical transition

We analyze the bifurcation diagrams of the coevolution under replicator dynamics and under knowledge feedback, and identify the critical points and thresholds that determine the system's fate. Our study elucidates that within the framework of coevolution under replicator dynamics, the critical line, defined as $x_0 = \dfrac{R_0 - (\mathbf{s}_{D^-})_R}{(\mathbf{s}_{C^-})_R - (\mathbf{s}_{D^-})_R}$, in the phase space constituted by the initial resource $R_0$ and the fraction of cooperators $x_0$, serves as a boundary separating two distinct stability regimes (see Fig. 7). The initial condition $\{R_0, x_0\}$ determines whether the system converges to the unsustainable or the sustainable solution. Our findings reveal that the system exhibits bi-stability and undergoes a critical transition between the two solutions near the critical value that corresponds to the critical line. We further observe a decrease in the resilience of the sustainable solution and its probability of occurrence with an increase in $\hat{e}_D$ or $A$. In contrast, for coevolution under knowledge feedback, we find that the system has higher resilience than coevolution under replicator dynamics, thereby underscoring the role of knowledge feedback in enhancing the resilience of the coevolving system.

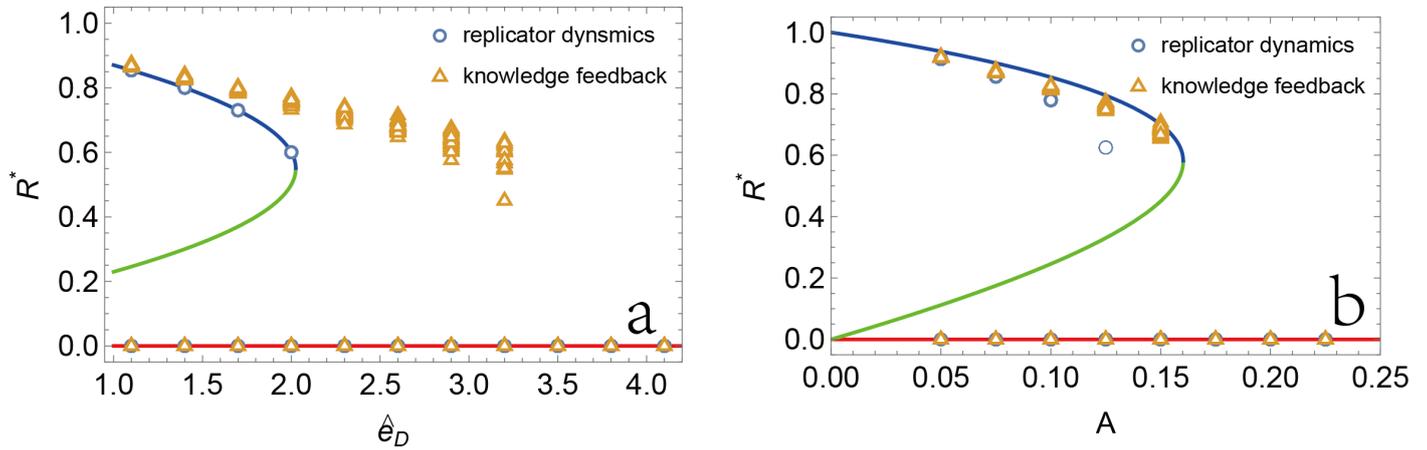

**Fig. 7 | Equilibrium states as a function of $\hat{e}_D$ and $A$.** (a) Bifurcation diagram of the equilibrium states of $R^*$ as a function of $\hat{e}_D$ where $A = 0.1$. (b) Bifurcation diagram of the equilibrium states of $R^*$ as a function of $A$ where $\hat{e}_D = 1.5$. The open markers represent numerical simulation results for coevolution with replicator dynamics and knowledge feedback. The blue, green, red line represent the stable sustainable state, the unstable state, and the stable unsustainable state from analytical derivation from coevolution with replicator dynamics, respectively. Parameters in these games are population size $N = 200$, growth rate $T = 2$, normalized extraction parameter $\hat{e}_C = 0.5$ and random initial condition $R_0, x_0$. We performed $n = 50$ independent numerical simulations for each case.

# Discussion

This paper advances the understanding of HES and CPR management by incorporating the Allee effect, a biological phenomenon that has been neglected in previous research. The Allee effect, which occurs when the per capita growth rate of a population decreases as the population size decreases, is a key characteristic of many biological resources, such as fisheries, forests, and pollinators, and has crucial implications for their management and conservation. By applying the Allee effect to CPR management within HES, our research introduces two coevolutionary models of resource and strategy under replicator dynamics and knowledge feedback. These models encapsulate various facets of resource dynamics and the players' behavior, such as the resource growth function, the extraction rates, and the strategy update rules.

The paper also connects to the evolutionary game theory literature by employing replicator dynamics to model strategy evolution and by examining the stability and bifurcations of the system. We adopt a general framework that couples two ODEs describing the temporal evolution of resource and players' strategies, and generalize it by incorporating different forms of the Allee effect and different behavioral rules and mechanisms for the players. We demonstrate that the Allee effect can cause bi-stability and critical transition in the system, resulting in either sustainable or unsustainable outcomes depending on the initial condition and the parameter configuration.

The paper has significant implications for the management and policy of CPR, as it reveals the potential challenges and opportunities for achieving a sustainable outcome in the presence of an Allee effect. We show that the Allee effect can enhance or hinder the resource sustainability, depending on the parameter values and the initial conditions. Therefore, the management and policy of CPR should take into account the Allee effect and its impact on the system dynamics and outcomes. Comparing the coevolution under knowledge feedback and that under replicator dynamics, we find that, regardless of the value of the extraction rate of cooperators, the sustainable region of knowledge feedback is consistently larger than that under replicator dynamics. Furthermore, when the systems are in a bi-stable state for the same parameters for both cases, the basin of attraction of the sustainable equilibrium under knowledge feedback is consistently larger than that under replicator dynamics. These results suggest that knowledge feedback enhances the resilience and sustainability of the coevolving system. Therefore, we imply that providing players with accurate and timely information about the resource and its Allee effect, and encouraging them to adapt their behaviors accordingly, can be an effective way to prevent the collapse of CPRs. We finally recommend that the management and policy of CPRs should be adaptive and responsive to the changes in the resource and the player behaviors, as the system can undergo critical transitions and exhibit multiple equilibria and path dependence.

However, the paper also has some limitations and assumptions that could be addressed or relaxed in future research. Firstly, we assume that the players have perfect information and rationality, which may not reflect many real-world situations. Future research could investigate how bounded rationality, learning, or uncertainty affect the outcomes. Secondly, we focus on a single resource and a homogeneous population of players, which may not represent the heterogeneity and interdependence of many CPRs. Future research could generalize the models to include multiple resources, multiple types of players, or network effects. Thirdly, we assume a homogeneous population of players with the same extraction rates and payoff functions. Future research could explore how heterogeneity among players, such as different preferences, beliefs, or learning abilities, would influence the system dynamics and outcomes. Lastly, we do not consider the effects of spatial structure, network topology, or social interactions on the system. Future research could study how the spatial distribution of the resource and the players, the connectivity and clustering of the players, or the communication and coordination among the players, would affect the evolution of cooperation and resource sustainability.

# Conclusion

In this paper, we explore the effects of the Allee effect, relationship between the resource availability and growth rate, on the dynamics and sustainability of CPRs. We employ two evolutionary game models to describe the interaction between the resource availability and the players' behavior, and analyze the stability and sustainability of the system under different scenarios. We reveal that the Allee effect can induce bi-stability and critical transition in the system, leading to either sustainable or unsustainable outcomes depending on the initial condition and the parameter configuration. We demonstrate that knowledge feedback enhances the resilience and sustainability of the coevolving system, and these results advances the understanding of HES and CPR management by incorporating the Allee effect. This paper contributes to the understanding of the complex and diverse dynamics of HES and significant implications for the management and policy of CPRs.